\documentclass[preprints,conferenceproceedings,accept,oneauthor,pdftex,10pt,a4paper]{Definitions/mdpi}

\pdfoutput=1

\firstpage{1}
\pubvolume{xx}
\issuenum{1}
\articlenumber{5}
\pubyear{2018}
\copyrightyear{2018}
\history{}

\usepackage[english]{babel}
\usepackage[utf8]{inputenc}
\usepackage{amssymb}
\usepackage{siunitx}
  \addto\extrasenglish{\sisetup{locale = UK, detect-all}}
\usepackage{subfig}
\usepackage{xspace}

\newcommand{\cel}{\ensuremath{\mathit{c}}}
\newcommand{\abs}[1]{\left\lvert#1\right\rvert}

\newcommand{\spp}{\ensuremath{\sqrt{s}}\xspace}
\newcommand{\sNN}{\ensuremath{\sqrt{s_{\text{NN}}}}\xspace}

\newcommand{\pp}{\text{pp}\xspace}

\newcommand{\pPb}{\text{p--Pb}\xspace}

\newcommand{\PbPb}{\text{Pb--Pb}\xspace}

\newcommand{\sigLong}{\ensuremath{\sigma_{\text{long}}^\text{2}}\xspace}
\newcommand{\pT}{\ensuremath{p_{\text{T}}}\xspace}
\newcommand{\pTiso}{\ensuremath{\pT^{\text{iso}}}\xspace}

\newcommand{\ET}{\ensuremath{E_{\text{T}}}\xspace}

\newcommand{\Rcone}{\ensuremath{R_{\text{cone}}}\xspace}

\newcommand{\gaminc}{\ensuremath{\gamma_{\text{inc}}}\xspace}

\newcommand{\gamdir}{\ensuremath{\gamma_{\text{dir}}}\xspace}

\newcommand{\gamdec}{\ensuremath{\gamma_{\text{dec}}}\xspace}

\newcommand{\Deta}{\ensuremath{\Delta\eta}\xspace}
\newcommand{\Dphi}{\ensuremath{\Delta\varphi}\xspace}

\newcommand{\piZ}{\ensuremath{\pi^{\text{0}}}\xspace}

\newcommand{\Rgam}{\ensuremath{R_{\gamma}}\xspace}

\Title{Direct Photon Measurements with the ALICE Experiment at the LHC$^{\dagger}$}

\Author{Erwann Masson $^{1}$\orcidA{}, on behalf of the ALICE Collaboration}

\AuthorNames{Erwann Masson}

\address{%
$^{1}$ \quad SUBATECH, IMT Atlantique, Université de Nantes, CNRS-IN2P3, Nantes, France; erwann.masson@subatech.in2p3.fr}

\corres{Correspondence: erwann.masson@subatech.in2p3.fr; Tel.: +33-251-858-669}

\firstnote{Presented at Hot Quarks 2018 - Workshop for young scientists on the physics of ultrarelativistic nucleus-nucleus collisions, Texel, The Netherlands, September 7-14 2018}

\abstract{In high energy hadron collisions, direct photons can be produced in various processes and are of particular interest to study the hot QCD medium since they escape it without being affected. In these proceedings are presented the latest ALICE experiment results concerning direct photon production in proton-proton (\pp), proton-lead (\pPb) and lead-lead (\PbPb) collisions. All measurements agree with pQCD calculations at high transverse momentum (\pT) and show no direct photon excess at low \pT in small systems while a low \pT signal is found in central \PbPb collisions.}

\keyword{direct photons; thermal radiation; isolated photons; pQCD}

\begin{document}

\section{Introduction}
\label{sec:intro}

The inclusive production of photons in hadron collisions (denoted by \gaminc) can be classified in two types: whereas decay photons (\gamdec) are emitted by decaying hadrons, mainly \piZ mesons through their $\gamma\gamma$ channel, the rest is called direct photons (\gamdir) so that $\gamdir = \gaminc-\gamdec$.

For $\pT\gtrsim\SI{5}{\giga\electronvolt}/\cel$ the direct photon production is dominated by $2\rightarrow2$ prompt processes involving incoming partons (Compton scattering and annihilation) as well as parton fragmentation, both well described by perturbative QCD (pQCD) at Next-to-Leading Order (NLO) and therefore suited for testing this theory. The fragmentation contribution, however, provides a partial information on the initial hard scattering but can be largely reduced thanks to an isolation method giving access to Leading Order (LO) photons and a way to constrain parton distribution functions (PDFs in \pp and nPDFs in heavy-ion collisions) \cite{jetphoxIchou}.

At lower \pT values the direct photon spectrum is mainly fed by thermalisation of the hadron gas following collisions and to a larger extent of the Quark-Gluon Plasma (QGP) expected in ultrarelativistic heavy-ion systems \cite{thermal_prod}. The production of these thermal photons strongly depends on the hot medium properties, thus they carry information on its space expansion and temperature which are valuable to explore the hadron matter phase diagram \cite{paquet_spacetime}.

The ALICE photon reconstruction techniques are introduced in Section \ref{sec:ALICE}, the latest results on low \pT direct photons are presented in Section \ref{sec:lowpT} and the high \pT isolated photon differential cross section measurement is discussed in Section \ref{sec:highpT}.

\section{Reconstructing photons with ALICE}
\label{sec:ALICE}

In the ALICE experiment \cite{ALICEexpe,ALICEPerfReport2}, photons can be reconstructed either through energy deposit in calorimeters or from $\mathrm{e}^+\mathrm{e}^-$ pairs produced when they convert in detector material. During LHC Run 1 two calorimeters were installed: the PHOton Spectrometer (PHOS) and the ElectroMagnetic Calorimeter (EMCal). PHOS consists of lead tungstate crystal cells of size $\num{2.2}\times\SI{2.2}{\centi\meter\squared}$ covering $\abs{\eta}<\num{0.13}$ and $\SI{260}{\degree}<\varphi<\SI{320}{\degree}$ in total whereas EMCal is built with lead-scintillator sampling layered cells with $\Deta\times\Dphi=\num{0.0143}\times\num{0.0143}$ each, covering $\abs{\eta}<\num{0.7}$ and $\SI{80}{\degree}<\varphi<\SI{187}{\degree}$. Photons are measured in both detectors through electromagnetic showers induced in their material, and adjacent cells with deposited energy are grouped in clusters. Photon selection is mainly based on cluster properties (e.g.\ their elongation) and on a charged particle veto.

Reconstructing photons from their conversion in detector material is made possible with the Photon Conversion Method (PCM). This technique is based on $\mathrm{e}^+$ and $\mathrm{e}^-$ tracks measured by the ALICE innermost detectors, the Inner Tracking System (ITS) and the Time Projection Chamber (TPC), and paired to locate neutral particle secondary vertices $V^0$ where they were emitted. Photons are selected with criteria on vertex properties (e.g.\ their topology). The probability for photons to convert in the ALICE material saturates at $\sim\SI{9}{\percent}$ but the PCM technique allows to reconstruct them within the large acceptance of the TPC, i.e.\ $\abs{\eta}<\num{0.9}$ and a full azimuth.

\section{Direct photons at low \pT in \pp, \pPb and \PbPb collisions: the subtraction method}
\label{sec:lowpT}

The ALICE Collaboration has been investigating direct photon production at low \pT in \PbPb collisions at $\sNN=\SI{2.76}{\tera\electronvolt}$ \cite{gamdir_PbPb_276}, \pp collisions at $\spp=\SI{2.76}{\tera\electronvolt}$ and \SI{8}{\tera\electronvolt} \cite{gamdir_pp_276_8} and more recently \pPb collisions at $\sNN=\SI{5.02}{\tera\electronvolt}$. The direct photon signal is obtained by statistically subtracting the decay photon component (\gamdec) from the inclusive photon yield (\gaminc) as follows:

\begin{equation}
	\gamdir = \gaminc - \gamdec = \left(\text{1}-\frac{\gamdec}{\gaminc}\right)\gaminc = \left(\text{1}-\frac{\text{1}}{\Rgam}\right)\gaminc,
\end{equation}
where $\Rgam = \gaminc/\gamdec \equiv (\gaminc/\piZ_\text{param})\big/(\gamdec/\piZ_\text{param})$ is called ``direct photon double ratio''. Within this method the decay photon contribution (\gamdec) is computed using a cocktail simulation parametrised with measured mother particle spectra (e.g.\ \piZ, $\eta$, $\mathrm{K}^\pm$) or transverse mass scaling (e.g.\ $\eta'$, $\omega$), its main contribution being \piZ mesons providing $\sim\SI{90}{\percent}$ of all decay photons at $\pT\approx\SI{1}{\giga\electronvolt}/\cel$. The $\piZ_\text{param}$ term is obtained by parametrising the measured \piZ spectrum with different models \cite{pi0_eta_prod_pp8}. As both \gaminc and \piZ yields can be measured with the same reconstruction techniques presented in Section \ref{sec:ALICE} (PCM, PHOS, EMCal) the use of \Rgam has the advantage of partially or completely cancelling several systematic uncertainties. Furthermore, direct photons can be extracted using different reconstruction techniques independently and all measurements can be combined taking care of statistical and systematic uncertainty correlations.

\begin{figure}
  \centering
	\hfill
  \begin{minipage}[b]{.305\textwidth}
		\subfloat[]{%
			\includegraphics[width=\textwidth]{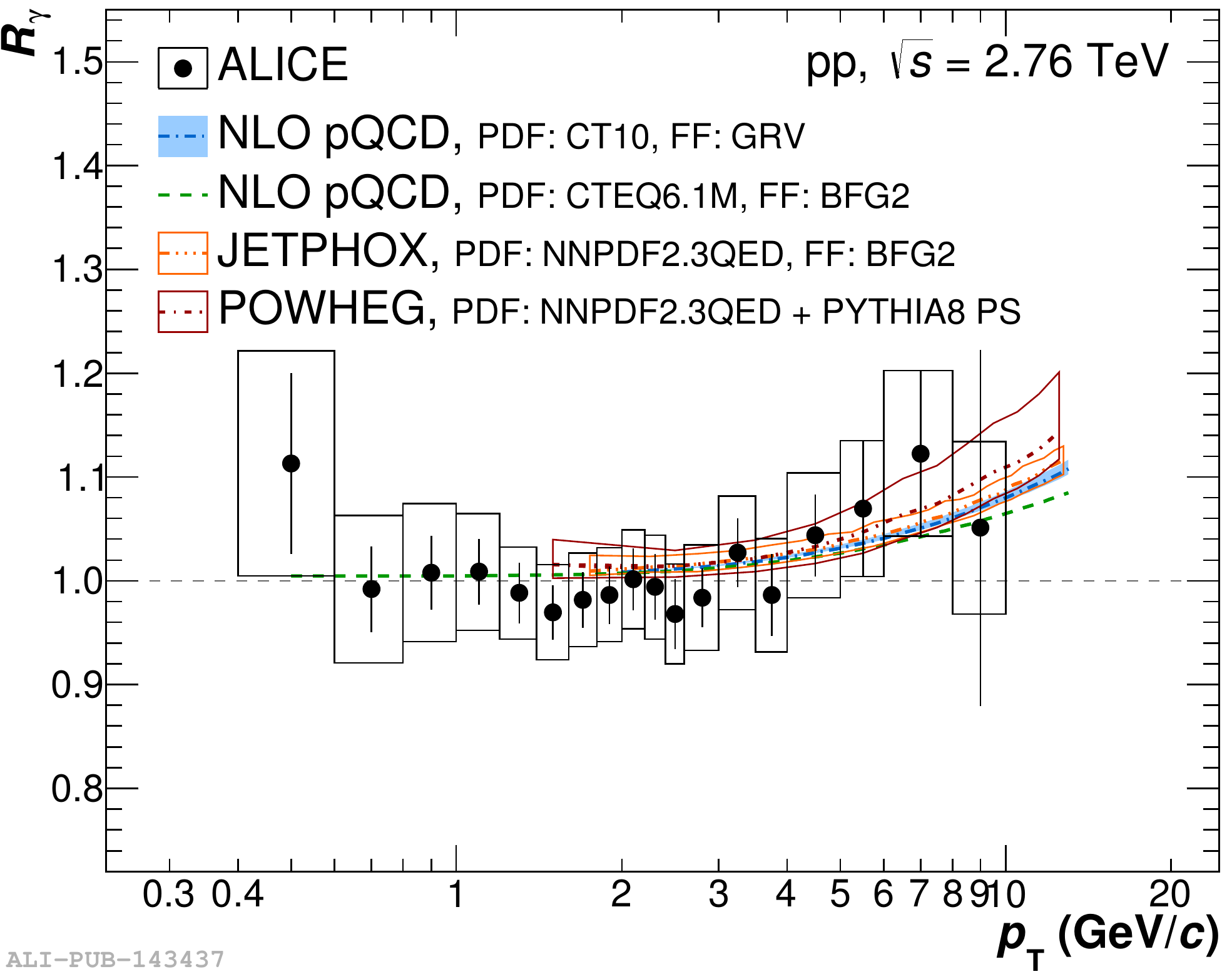}
			\label{sfig:Rgam_pp_276}
		} \\[-0.65\baselineskip]
		\subfloat[]{%
			\includegraphics[width=\textwidth]{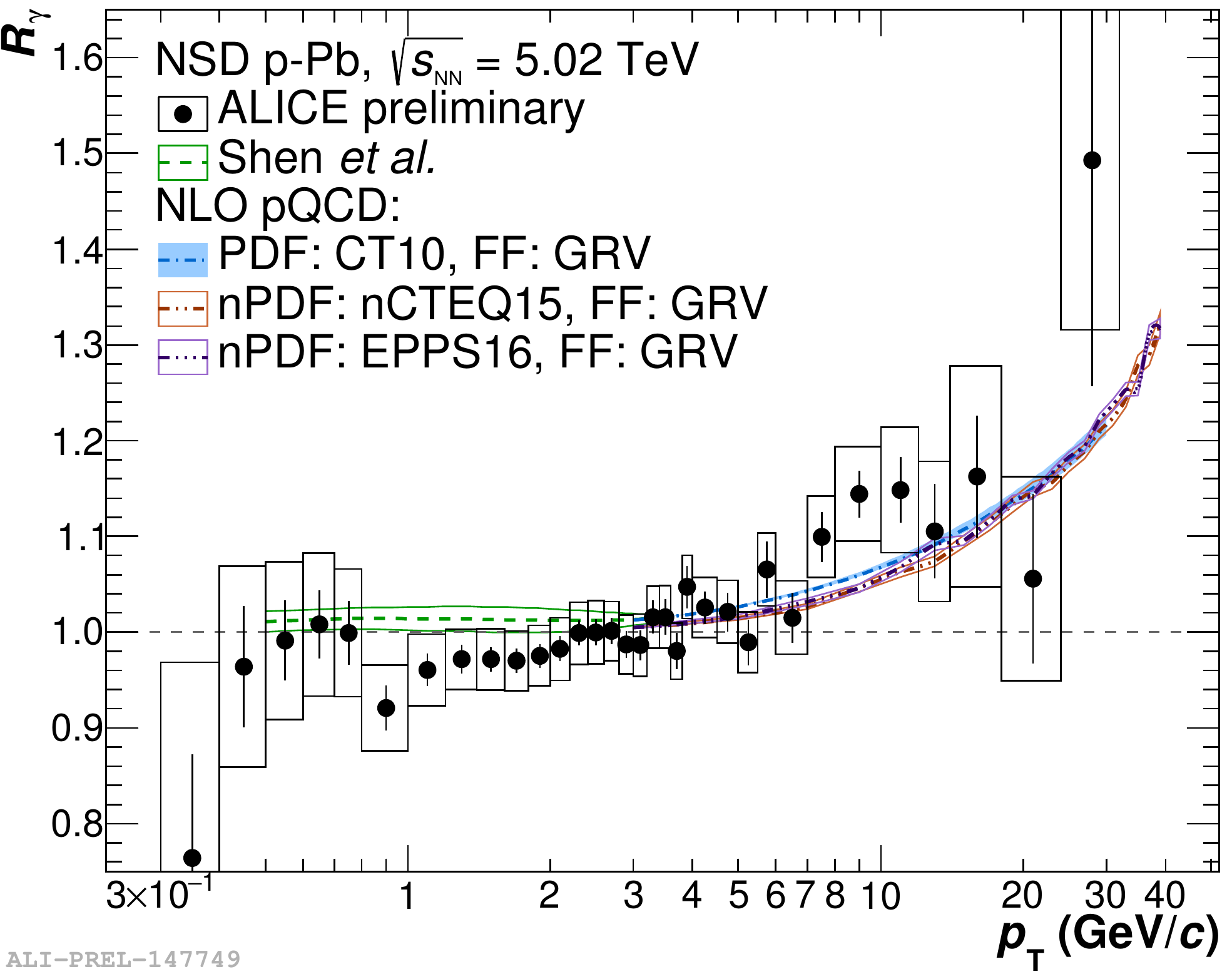}
			\label{sfig:Rgam_pPb_5}
		}
  \end{minipage}
	\hfill\hfill
	\begin{minipage}[b]{.458\textwidth}
		\subfloat[]{%
			\includegraphics[width=\textwidth]{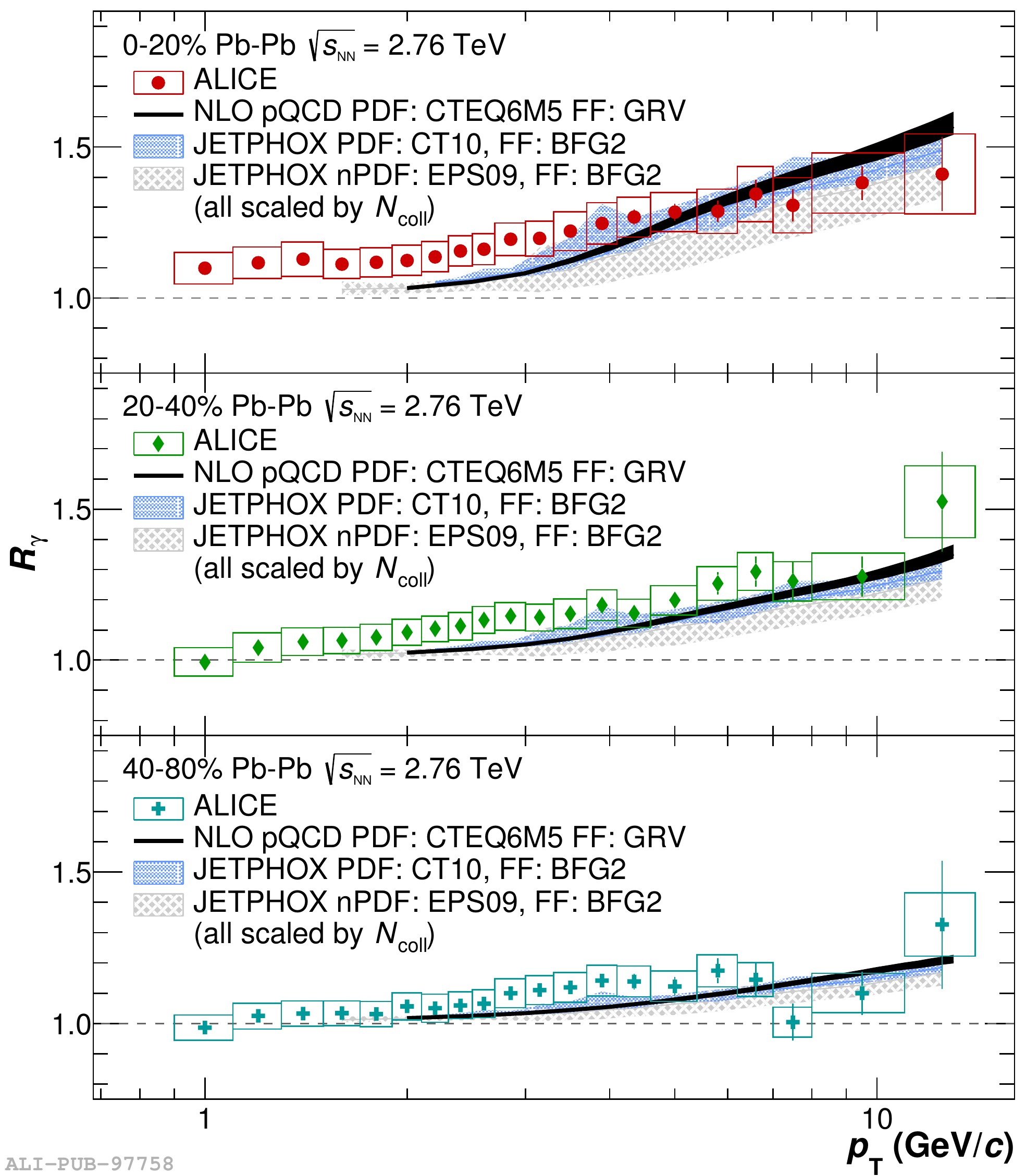}
			\label{sfig:Rgamma_PbPb_276}
		}
	\end{minipage}
	\hfill\hfill
  \caption{Direct photon double ratios measured in \textbf{\protect\subref{sfig:Rgam_pp_276}} \pp collisions at $\spp=\SI{2.76}{\tera\electronvolt}$ \cite{gamdir_pp_276_8}, \textbf{\protect\subref{sfig:Rgam_pPb_5}} \pPb collisions at $\sNN=\SI{5.02}{\tera\electronvolt}$ and \textbf{\protect\subref{sfig:Rgamma_PbPb_276}} \PbPb collisions at $\sNN=\SI{2.76}{\tera\electronvolt}$ \cite{gamdir_PbPb_276}. Both results are compared with theory including pQCD at NLO and hydrodynamic calculations (see \cite{gamdir_PbPb_276,gamdir_pp_276_8} for details).}
	\label{fig:Rgam}
\end{figure}

The double ratios measured by ALICE in different collision systems are shown in Figure \ref{fig:Rgam}. All results are compared with pQCD calculations at NLO (see \cite{gamdir_PbPb_276,gamdir_pp_276_8} for details). An agreement with theory is observed for $\pT\gtrsim\SI{5}{\giga\electronvolt}/\cel$ in all systems and centrality classes (for \PbPb) supporting the prompt photon production scenario. Whereas \Rgam is compatible with unity at very low \pT in \pp and \pPb collisions, a significant excess (\SIrange{10}{15}{\percent}) is observed in the most central \PbPb collisions indicating that another source of direct photons is present. It has also been shown \cite{gamdir_PbPb_276} that the resulting direct photon yield is compatible with several hydrodynamic models assuming the formation of a QGP and that the medium effective temperature measured by ALICE at the LHC is $\sim\SI{30}{\percent}$ higher than observed by PHENIX at the RHIC \cite{phenix_AuAu_200}, consistently with the expectations comparing collision energies.

\section{Direct photons at high \pT in \pp collisions: the isolation method}
\label{sec:highpT}

The direct photon high \pT domain has been explored in \pp collisions at $\spp=\SI{7}{\tera\electronvolt}$ by the ALICE Collaboration using the EMCal reconstruction technique. If photon contributions from LO and fragmentation processes cannot be disentangled at the detector level, one can strongly reduce the latter using an isolation method on a candidate photon sample. This technique consists in measuring the total activity \pTiso in a cone of radius $\Rcone=\sqrt{(\eta-\eta_\gamma)^\text{2}+(\varphi-\varphi_\gamma)^\text{2}}$ defined around a candidate photon located in $(\eta_\gamma,\varphi_\gamma)$, summing the transverse momentum of all particles located in $(\eta,\varphi)$. The neutral contribution to this activity is measured with EMCal neutral clusters whereas the charged contribution is determined with tracks from the ITS and TPC detectors. In the work presented here, photons are considered isolated for $\Rcone=\num{0.4}$ and $\pTiso<\SI{2}{\giga\electronvolt}/\cel$, values determined with pQCD calculations to reduce the fragmentation contribution from \SI{45}{\percent} to \SI{15}{\percent} of all direct photons at $\ET=\SI{60}{\giga\electronvolt}$ \cite{jetphoxIchou}.

Another key parameter to select direct photons is the EMCal cluster elongation mentioned in Section \ref{sec:ALICE}. It is denoted by \sigLong and used to discriminate decay photons (mainly from $\piZ\rightarrow\gamma\gamma$ inducing elongated clusters with high \sigLong values) and direct photons (circular clusters with low \sigLong values), and therefore to reduce the former contribution. However, the isolated and circular cluster sample still contains a residual contamination of clusters induced, for instance, by single photons from asymmetric \piZ decays. This contamination --- in other words, the isolated photon sample purity --- is estimated using a double sideband method in the $\big(\pTiso,\sigLong\big)$ phase space, assuming no correlation between these parameters and a negligible signal leakage in background bands \cite{ATLAS_abcd}. These strong hypotheses are carefully investigated and corrected with Monte Carlo simulations.

\begin{figure}
  \centering
	\hfill
	\subfloat[]{%
		\includegraphics[width=0.45\textwidth]{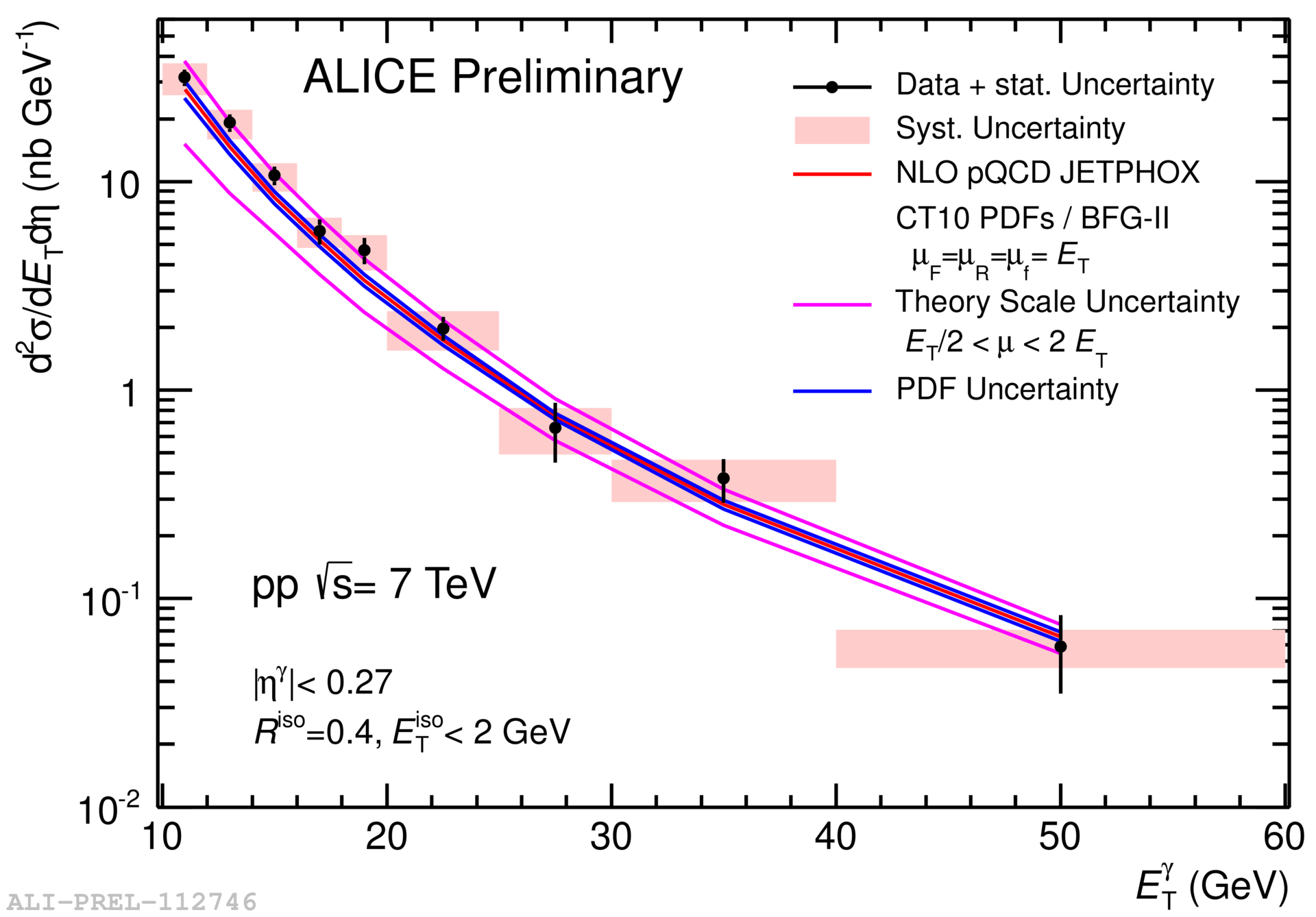}
		\label{sfig:Xsec_pp_7}
	}
	\hfill
	\subfloat[]{%
		\includegraphics[width=0.45\textwidth]{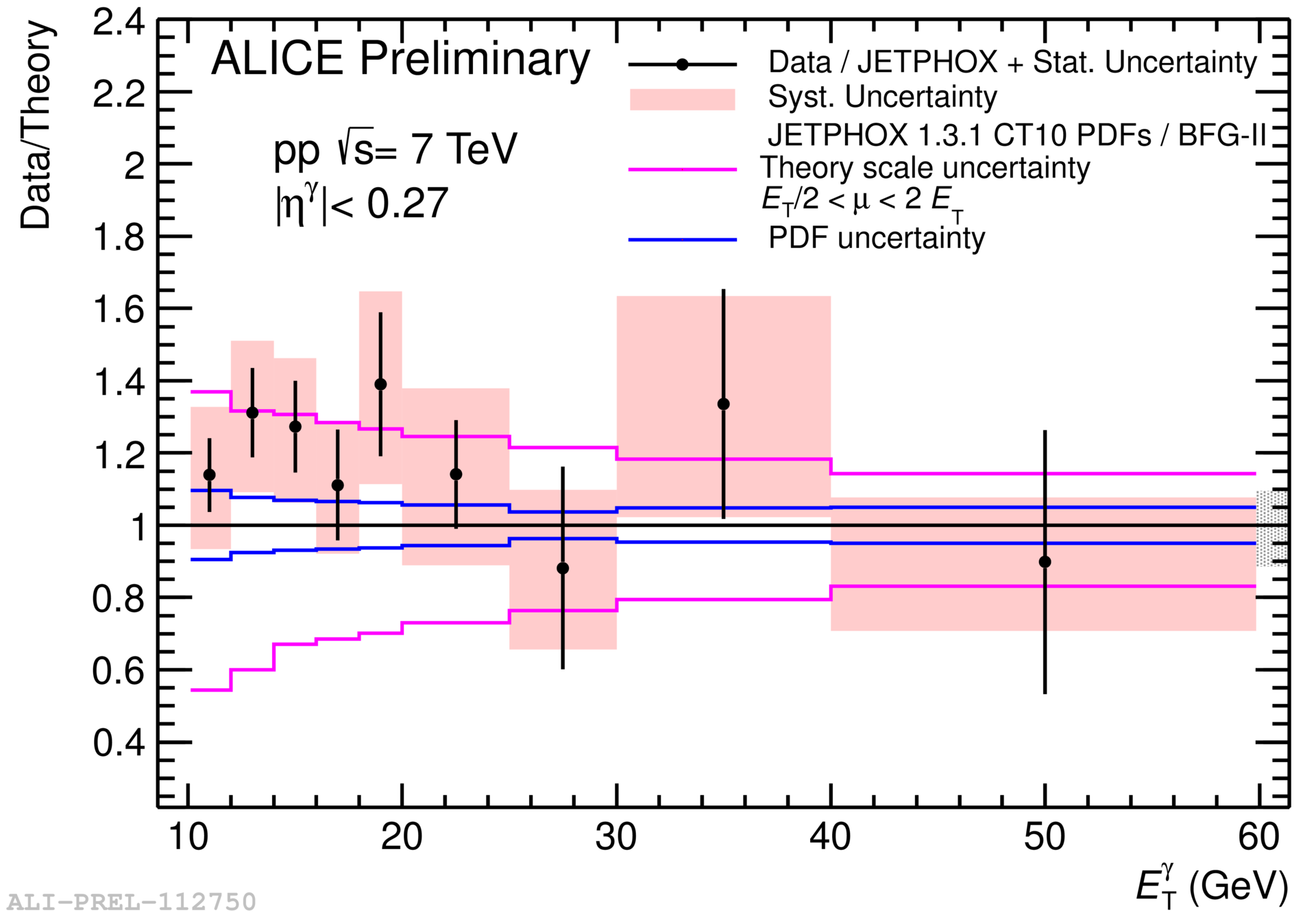}
		\label{sfig:ratio_pp_7}
	}
	\hfill\hfill
  \caption{Isolated photon differential cross section measured in \pp collisions at $\spp=\SI{7}{\tera\electronvolt}$, \textbf{\protect\subref{sfig:Xsec_pp_7}} compared with pQCD calculations at NLO \cite{jetphox} and \textbf{\protect\subref{sfig:ratio_pp_7}} divided by these calculations.}
	\label{fig:Xsec}
\end{figure}

The cluster raw yield is corrected by this purity and the efficiency accounting for reconstruction, identification and isolation steps, then scaled by the integrated luminosity associated to the measurement. The resulting isolated photon differential cross section in \pp collisions at $\spp=\SI{7}{\tera\electronvolt}$ is shown in Figure \ref{fig:Xsec} as well as pQCD calculations at NLO \cite{jetphox} for comparison. A reasonable agreement is observed between data and theory from \SI{10}{\giga\electronvolt} to \SI{60}{\giga\electronvolt} and this measurement extends the \ET reach down compared to previous results published by the ATLAS \cite{ATLAS_abcd} and CMS \cite{CMS_pp} collaborations.

\section{Summary}

The latest ALICE experiment results about direct photon production in different collision systems were presented in these proceedings. Two methods are used to extract the signal of interest depending on the photon momentum. At low \pT, decay photons are statistically subtracted from the inclusive photon yield to obtain the direct photon contribution using several independent reconstruction techniques. This measurement in \pp and \pPb collisions is fully consistent with pQCD calculations at NLO whereas in central \PbPb collisions an excess compatible with the presence of QGP is observed. At high \pT, the direct photon yield is extracted with an isolation method allowing to strongly reduce the fragmentation and decay photon contributions. This measurement in \pp collisions extends the \pT reach compared to previous results while showing a good agreement with pQCD throughout the probed range. Other collision systems and energies are being investigated currently to get a comprehensive picture of the direct photon production.

\vspace{6pt}

\reftitle{References}
\externalbibliography{yes}
\bibliographystyle{Definitions/mdpi.bst}
\bibliography{2018_HQ_Proceedings_ErwannMasson_biblio.bib}


\end{document}